\documentclass[epj]{svjour}

\usepackage{graphicx}
\usepackage{fancyhdr,url}
\usepackage{amsmath,amssymb}
\def\makeheadbox{{
 }}
\def\mytitle{My title} 
\def\myauthors{My name}  
\def\mytype{My type of session}
\def\mysession{My session}

\def\mytitle{Multiparticle SUSY simulations at LHC \& ILC} 
\def\myauthors{J\"urgen Reuter}    
\def\mytype{Parallel Talk}    
\def\mysession{Colliders - SUSY Phenomenology}

\begin{document}
\title{Multiparticle SUSY simulations at LHC \& ILC: Off-Shell effects,
  interferences and radiative corrections} 
\author{J\"urgen Reuter
\thanks{\emph{Email:} juergen.reuter@desy.de}%
}
\institute{University of Freiburg, Institute
 of Physics, Germany
}
%


\date{}
\abstract{
The interesting but difficult phenomenology of supersymmetric models
at the LHC and ILC demands a corresponding complexity and maturity from
simulation tools. This includes multi-particle final states, reducible
and irreducible backgrounds, spin correlations, real emission of
photons and gluons, virtual corrections etc. Most of these topics are
included in the multi-particle Monte Carlo (MC) Event generators
Madgraph, WHIZARD and Sherpa. A comparison of these codes is shown,
with a special focus on the new release of WHIZARD. I show 
examples for the necessity of considering full matrix elements  
with all off-shell effects and interferences for multi-particle final
states in supersymmetric models and give a status report on ongoing
projects for simulations of SUSY processes at the LHC with these
codes, including all of the abovementioned corrections.
} 
\maketitle
%

\newcommand{\rep}[1]{\mbox{\boldmath$#1$}}%
\newcommand{\arep}[1]{\mbox{\boldmath$\overline{#1}$}}%
\newcommand{\vev}[1]{{\langle #1 \rangle}}

\section{The need for multi-particle event generators}

Supersymmetry (SUSY) is the prime example for beyond the Standard
Model (SM) physics as a solution to the hierarchy problem. Compared to the SM,
it doubles the particle spectrum, and the SUSY-breaking terms induce a
vast number of new parameters. The analysis goal of the future
experiments at the LHC and an ILC include mass measurements to
determine the SUSY spectrum, the access of the new particles' spin
encoded in angular correlations, and finally, by means of coupling
measurements the proof that the new particles really fill
super-multiplets. Together with the complicated experimental set-up to
dig out the informations mentioned above, there is an utter need for
precise theoretical predictions of SUSY processes: first of all, to
access them from the SM background, secondly since SUSY processes are
themselves backgrounds for (more complicated) SUSY
processes. Ultimately, the goal is to extract the SUSY Lagrangian
parameters as precisely as possible to make predictions about the GUT
structure and the SUSY breaking mechanism~\cite{SPA}. 

\begin{figure}
\begin{center}
\includegraphics[width=.24\textwidth]{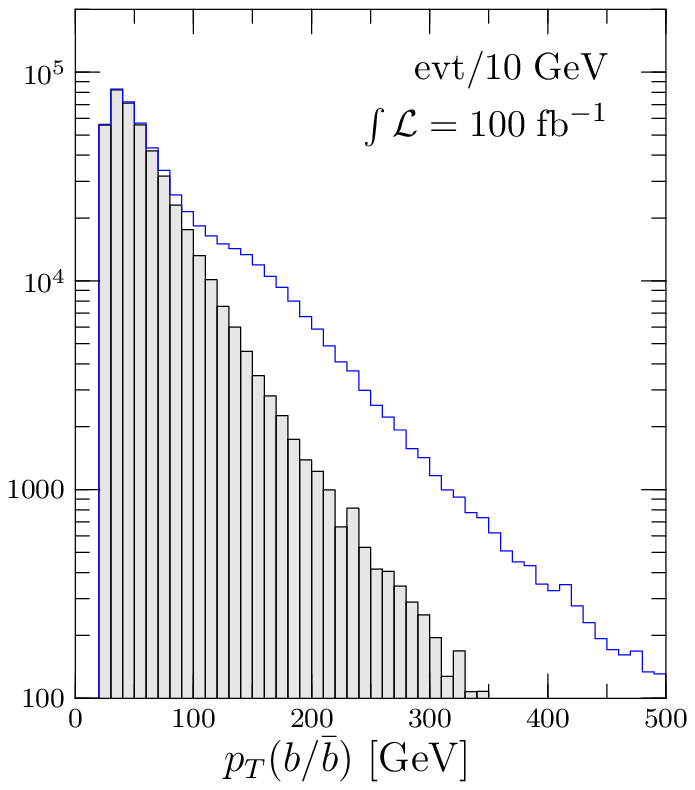}
\includegraphics[width=.24\textwidth]{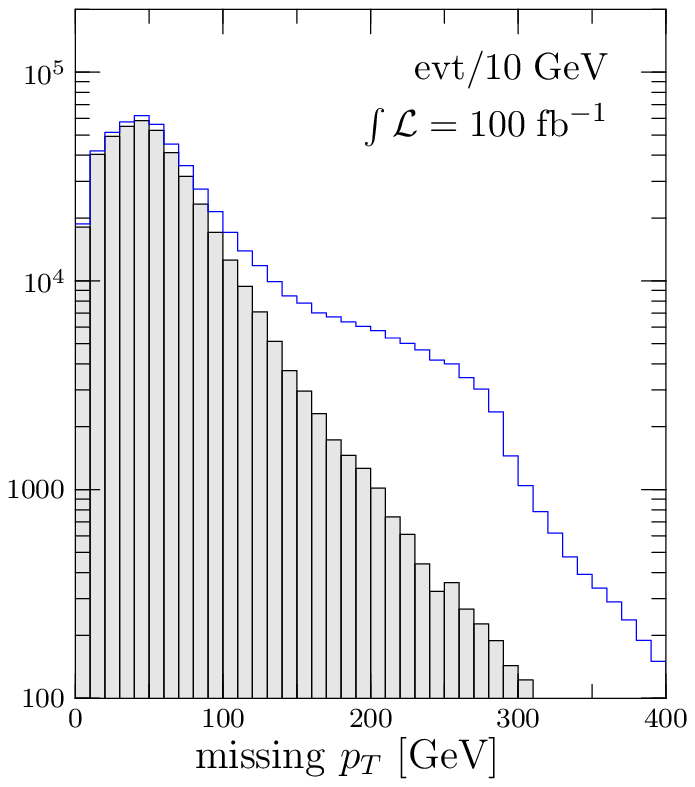}
\end{center}
\caption{$p_T(b)$ and missing $p_T$ distribution for the process $pp
  \to b \bar b\tilde{\chi}_1^0\tilde{\chi}_1^0$ resulting from sbottom
  pair production at the LHC. The SM background is in grey, the signal
  in blue.}
\label{fig:lhc_pt}
\end{figure}

In general, corrections to SUSY (or in general BSM) processes can be
grouped into six categories: 1) loop corrections to SUSY production
and decay processes, 2) real gluon/photon radiation, 3)
non-factorizable, maximally resonant photon exchange between
production and decay, 4) off-shell kinematics for the signal process,
5) irreducible background from all other SUSY processes, and 6)
reducible, but experimentally indistinguishable background from SM
processes. Topics 1) and 3) are dealt with in~\cite{trobens}. Here, we
will focus on the last three points. 

SUSY processes reach a very high level of complexity already at tree
level: e.g.~the process $e^+ e^- \to b\bar b e^+
e^-\tilde{\chi}_1^0\tilde{\chi}_1^0$ -- which is just
$\tilde{\chi}_2^0\tilde{\chi}_2^0$   pair production -- has $\sim
66,500$ Feynman diagrams. Several different production channels like
$\tilde{\chi}^0_i \tilde{\chi}^0_j$, $\tilde{b}_i\tilde{b}_j^*$ and
$\tilde{e}_i \tilde{e}_j^*$ interfere with each other and need to be
disentangled experimentally as well as in the simulation. One has to
add SM backgrounds like $e^+ e^- \to b\bar b e^+
e^- \nu\bar\nu$. And processes can be much more complicated at the
LHC, but also at the ILC. To make simulation of such complicated
multi-particle processes feasible, among the following three levels of
complexity usually one of the two first are used: first, there is the
narrow-width approximation (NWA), where intermediate states are
produced on-shell and multiplied by the corresponding branching
ratio. Secondly, there is the Breit-Wigner approximation (BWA), where
the on-shell state's width is accounted for by folding in a
Breit-Wigner propagator. Finally, there are the full matrix elements
including all intermediate off-shell states contributing to the same
exclusive final state. Programs like PYTHIA, HERWIG, SUSYGEN and
ISAJET use the first two approaches. Here, we will show that this is
not sufficient and that for BSM models full matrix elements have to
used.   


\section{SUSY simulations at the LHC}  

For simulations of multi-particle SUSY processes at the LHC, the
three packages WHIZARD~\cite{omega,whizard}, Sherpa \cite{sherpa} and 
Madgraph~\cite{madgraph} have been developed. All of these programs
follow the SUSY Les Houches accord (SLHA) \cite{SLHA}. Since the MSSM
is a fairly complicated model with several thousand different
vertices, it is mandatory to validate the correctness of its
implementation in the codes. This has been done in~\cite{catpiss} by
means of unitarity and gauge invariance checks as well as a direct
comparison of the three programs. To test all phenomenologically
relevant couplings, more than 500 different processes had to be
checked; they are listed here~\cite{catpiss,comparison}, which might
serve as a standard reference. Furthermore, supersymmetric Ward- and
Slavnov-Taylor identities have been checked~\cite{swi}. 

\begin{figure}
  \begin{center}
    \includegraphics[width=.24\textwidth]{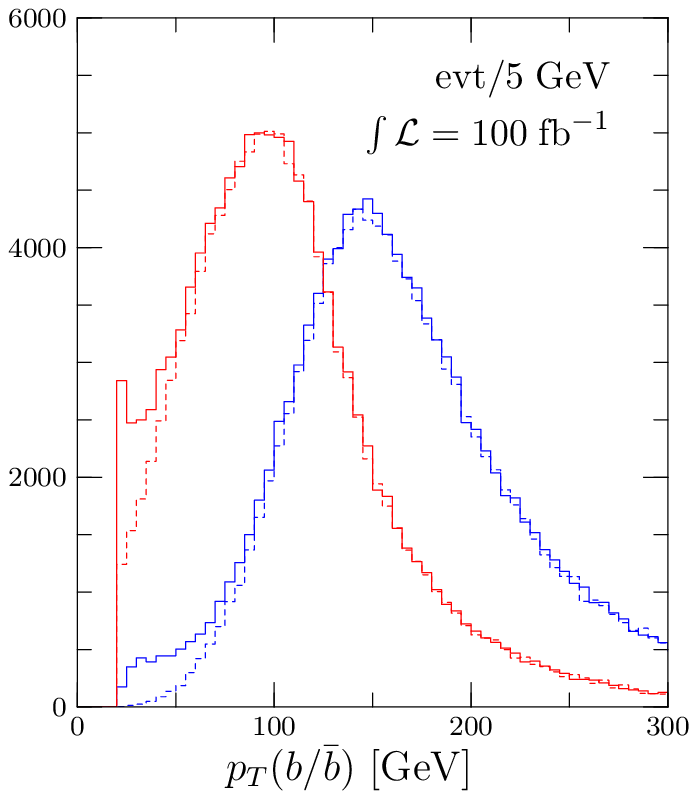}
    \includegraphics[width=.24\textwidth]{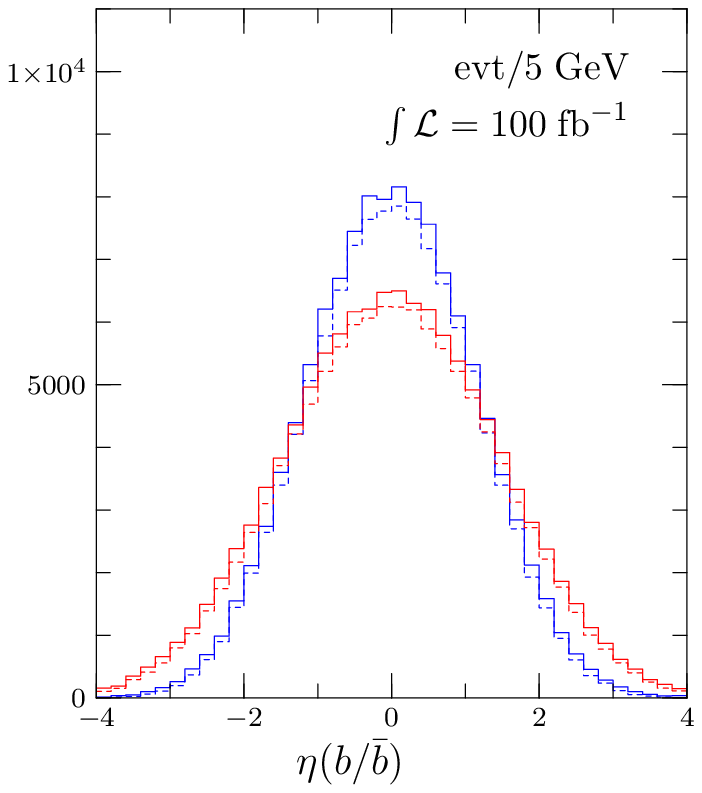}
  \end{center}
  \caption{Process: $pp\to b\bar b \tilde{\chi}_1^0 \tilde{\chi}_1^0$,
  distributions of the harder and softer $b$ jet in red
  and blue, respectively: on the left $p_T$, on the right in
  $\eta$. Full lines are full matrix elements, dashed ones
  Breit-Wigner approximation.}
  \label{fig:lhc_distribution}
\end{figure}

To study the influence of interferences and off-shell effects as well
as of radiative corrections, we chose a mSUGRA-inspired parameter
point with non-universal scalar masses (note that the following
discussion does by no means depend on that special point). The sbottom
masses are 295 and 400 GeV, respectively, the lightest Higgs is
directly above the LEP exclusion limit. It has a large (47\,\%)
branching ratio of invisible decays to the neutralino LSP. The first
generation squarks and sleptons are at 430 and 205 GeV,
respectively. The point is compatible with all low-energy constraints
like $b\to s\gamma$, $B_s \to \mu^+ \mu^-$, $\Delta\rho$, $g_\mu - 2$,
and cold dark matter. The focus here lies on sbottom production with a
BR$(\tilde{b}_1 \to b \tilde{\chi}_1^0) = 43.2\,\%$.

\newcommand{\sla}[1]{/\!\!\!#1}

\begin{figure}
  \begin{center}
    \includegraphics[width=.24\textwidth]{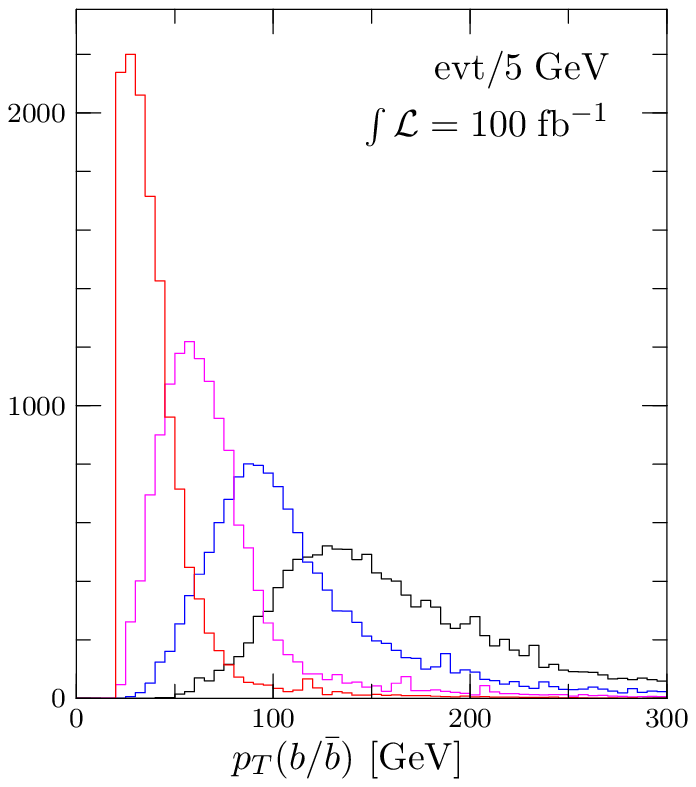}
    \includegraphics[width=.24\textwidth]{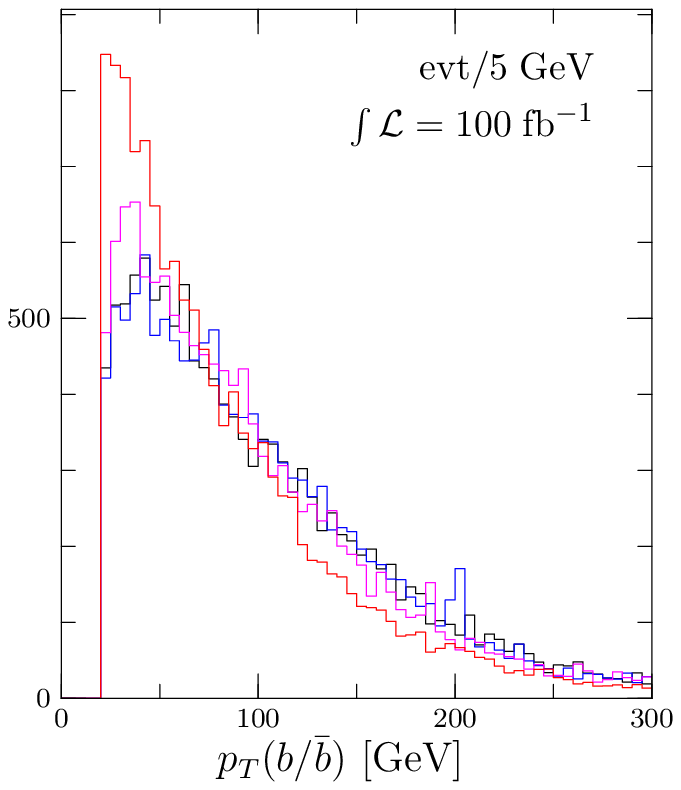}
  \end{center}
  \caption{$p_T$ distributions of the four $b$ jets in the process $pp
  \to b\bar bb\bar b \tilde{\chi}_1^0\tilde{\chi}_1^0$, on the left:
  ordered according to their $p_T$, on the right: ordered according to
  their centrality.}
  \label{fig:bbbb}
\end{figure}

Fig.~\ref{fig:lhc_pt} shows the parton-level distributions for $pp
\to b\bar b\tilde{\chi}_1^0\tilde{\chi}_1^0$, where we used standard
cuts of $p_{T,b} > 20$ GeV, $\eta_b < 4$, and $\Delta R_{bb} >
0.4$. The main SM background comes from $pp \to b\bar b \nu\bar\nu$
and is shown in gray. The signal is in blue, the $p_{T,b}$
distribution on the left, the $\sla{p}_T$ distribution on the
right. Here, the bumb coming from the sbottom production is clearly
visible. Generally, the signal jets are harder than the jets from the
SM background. In Fig.~\ref{fig:lhc_distribution}, the $p_{T,b}$ and $\eta_b$
distributions are plotted for the harder (red) and the softer (blue)
of the two jets. From the right plot we see that the harder jet is
much more central as expected from phase space restrictions. The full
lines here denote usage of full matrix elements, while the dashed
lines are the BWA. There is a clear discrepancy between the curves,
stemming mainly from $b\bar b Z^*$ diagrams; these discrepancies are
only in the low-$p_T$ region which is anyhow cut out. The reason why
off-shell decays are irrelevant here is the large mass splitting 
from 295 to 45 GeV between the sbottom and the LSP. For general SUSY
decay cascades this is not the case as will be explained in
section~\ref{sec:offshell}.

\begin{figure}
  \begin{center}
    \includegraphics[width=.24\textwidth]{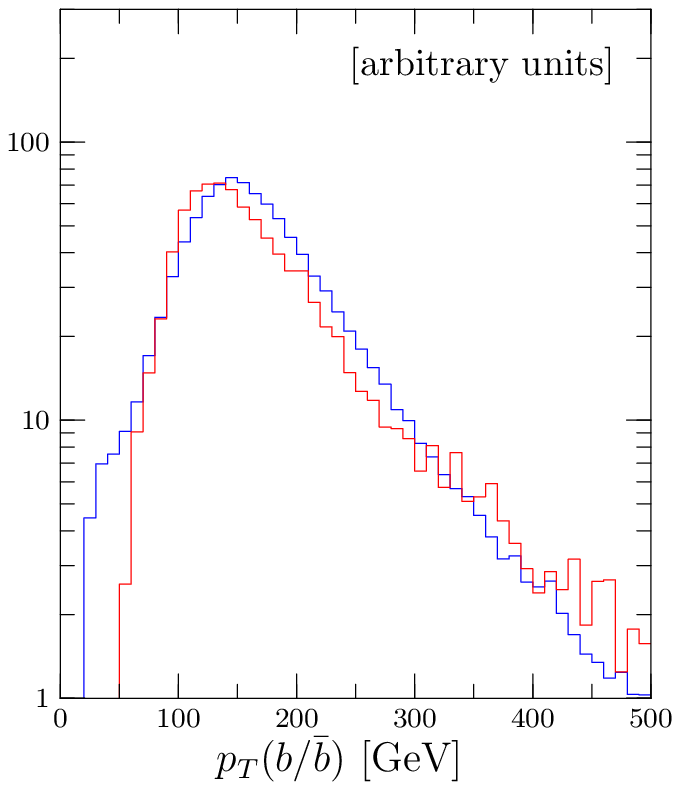}
    \includegraphics[width=.24\textwidth]{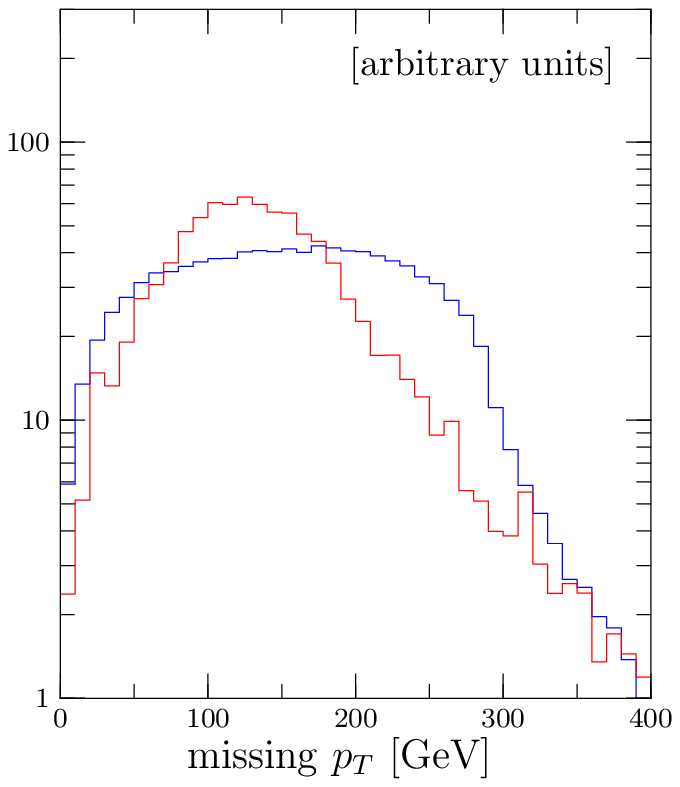}
  \end{center}
  \caption{$p_{T,b}$ and $\sla{p}_T$ distributions for the processes
  with two $b$ jets in blue and with four $b$ jets in red.}
  \label{fig:lhc_comparison}
\end{figure}

As a next step, we studied the influence of real corrections on the
detection of the signal, i.e. bottom-jet radiation from the initial
state ($g\to b\bar b$ splitting) as combinatorial background. Using
the above cuts for the $b$ jets, the perturbation series is
well-behaved, the cross section ceases from 1177 fb to 130.7 fb when
including the two bottom jets. The process with four $b$ jets, $pp \to
b\bar bb\bar b \tilde{\chi}_1^0 \tilde{\chi}_1^0$, is at the border of
feasibility for SUSY LHC simulations with 32,000 diagrams, 22 color
flows and several thousand phase space channels. The simulation has
been performed with WHIZARD, which is well-suited for physics beyond
the SM~\cite{omwhiz_bsm}. Fig.~\ref{fig:bbbb} shows the $p_T$ 
distributions of the four $b$ jets, ordered according to their $p_T$
on the left and to their pseudorapidity on the
right. The right figure shows that only the most forward jet is
considerably softer, and that a forward discrimination between 
ISR and signal (decay) jets can be quite
intricate. Fig.~\ref{fig:lhc_comparison}, on the other hand, shows
that the jet structure (left hand side) itself does not change
considerably from the case with two (blue) and four $b$ jets
(red). Since more hard jets have to be produced, the PDFs decrease
the maximum a little bit. The missing $p_T$ is shifted to lower 
values because light particles balance out the event.


\section{SUSY simulations at the ILC}

\begin{figure}
  \begin{center}
    \includegraphics[width=.42\textwidth]{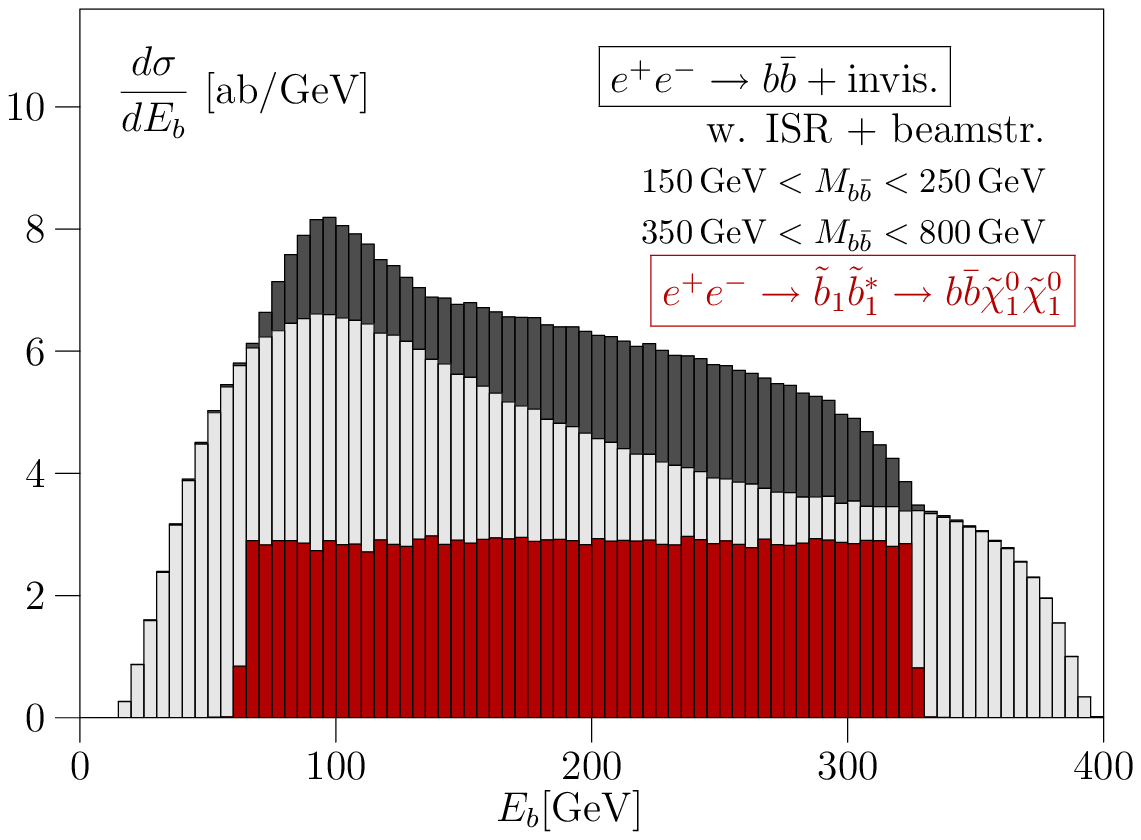}

    \vspace{2mm}

    \includegraphics[width=.42\textwidth]{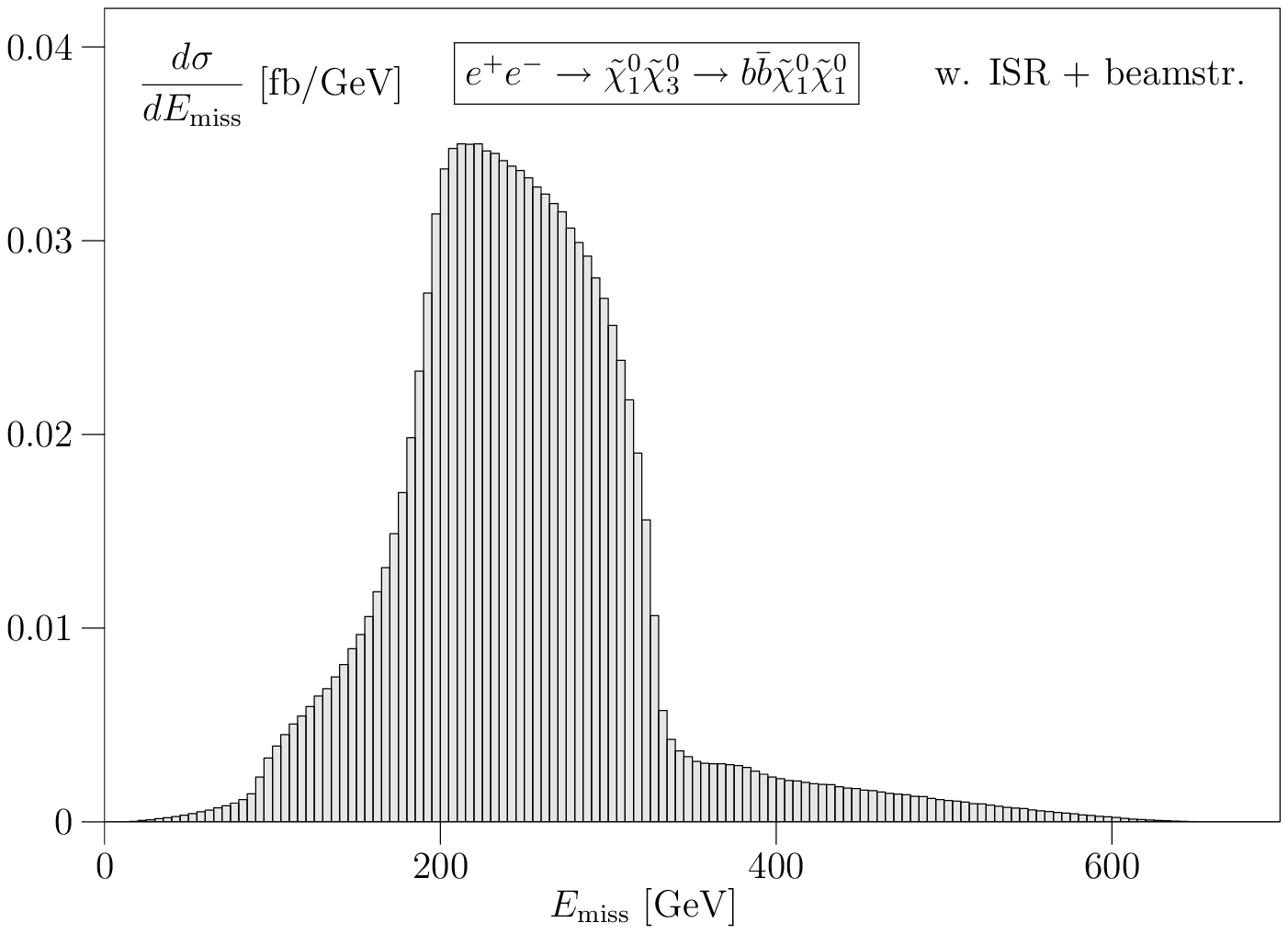}
  \end{center}
  \caption{Top: energy of the $b$ jet in the process $e^+ e^- \to b \bar b
  + \sla{E}$ at the ILC. In red the NWA, in light gray the SM 
  background, and in dark gray the signal with full matrix elements on
  top of it. ISR and beamstrahlung are included. Bottom: same, but
  $\sla{E}$ distribution.}
  \label{fig:ilc_box} 
\end{figure}

At the ILC, we consider the same final state, $e^+ e^- \to b\bar
b \tilde{\chi}_1^0\tilde{\chi}_1^0$. Here, the production is
electroweak, whence a lot more of intermediate states contribute:
$Zh$, $ZH$, $HA$, $\tilde{\chi}_1^0\tilde{\chi}_2^0$,
$\tilde{\chi}_1^0\tilde{\chi}_3^0$,
$\tilde{\chi}_1^0\tilde{\chi}_4^0$, $\tilde{b}_1\tilde{b}_1^*$,
$\tilde{b}_1\tilde{b}_2^*$. Especially, the background from heavy
Higgses and the heavier neutralinos is quite severe. The irreducible
SM background is mainly from $WW$ fusion: $e^+e^- \to b\bar b
\nu\bar\nu$. To extract the sbottom from the SUSY backgrounds, one has
to cut out the regions in the $M_{b\bar b}$ spectrum from 150 to 250
GeV as well as from 350 to 800 GeV. Applying these cuts, the result
for the cross section of the signal is $0.487$ fb ($0.375$ fb with ISR
and beamstrahlung); the NWA gives 2.314 fb, which is one order of
magnitude away from the full result. The reason is that interferences
with heavy neutralino three-body decays affect the decay kinematics of
the sbottoms. Hence, for ILC it is absolutely mandatory to use full
matrix elements and multi-particle event generators for SUSY
simulations.   


\section{Origin of off-shell effects}
\label{sec:offshell}

In this section, we briefly explain the reason for the appearance of
large off-shell and interference effects for SUSY processes at the
LHC (for a clear derivation of these effects,
see~\cite{berdine}). Although, here we focus on SUSY, all new physics
scenarios which have a conserved matter parity (responsible for dark
matter) and a complicated spectrum without too large hierarchies in
common, share the same features. Their phenomenology grossly consists
of decay chains with quasi-degeneracies among mother and daughter
particles. Including the full momentum dependence of the decay matrix 
elements affects the momentum dependence of the propagators:
performing phase space integrals does not result in simple
Breit-Wigner terms that factor out. Although naively, the importance
of off-shell effects is expected to scale with the width to mass ratio
$\Gamma/M$ for the resonances, these factors can be largely enhanced
by powers of the (small) inverse velocity of the daughters,
$1/\beta^n$, coming from the near-degeneracies of mothers and
daughters. Furthermore, interferences with diagrams where only one of
the two decay chains is resonant can be quite large. 

Altogether, these effects lead to drastic deviations from the NWA:
effective branching ratios (which are measured via their exclusive
final states) are shifted by $20-40$ \%. Charge and chirality
asymmetries could appear in the decays of gluinos:
e.g.~$\tilde{g}\tilde{g} \to b\tilde{b}_1^* b\tilde{b}_1^*$ vs. $\bar
b \tilde{b}_1 \bar b \tilde{b}_1^*$ could have different 
effective branching ratios since they have different subsequent decay
chains which have different interferences with non-reso\-nant diagrams. 
A chirality asymmetry is a non-equality between the two decays
$\tilde{g} \tilde{g} \to \tilde{q}_L \tilde{q}_L jj$ vs. $\tilde{q}_R
\tilde{q}_R jj$ for the same reasons as
above. 

\begin{figure}
  \begin{center}
    \includegraphics[width=.24\textwidth]{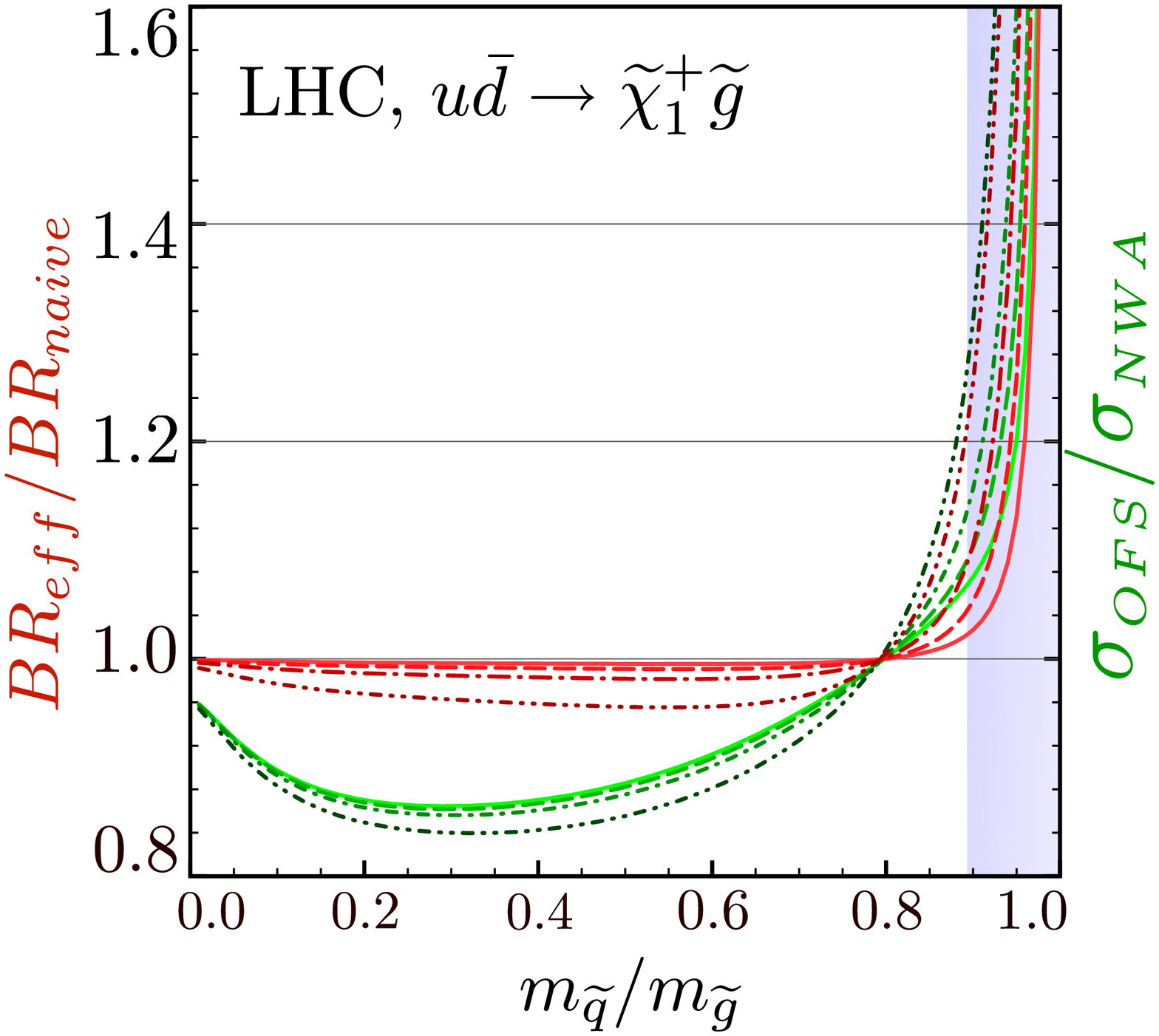}
    \includegraphics[width=.24\textwidth]{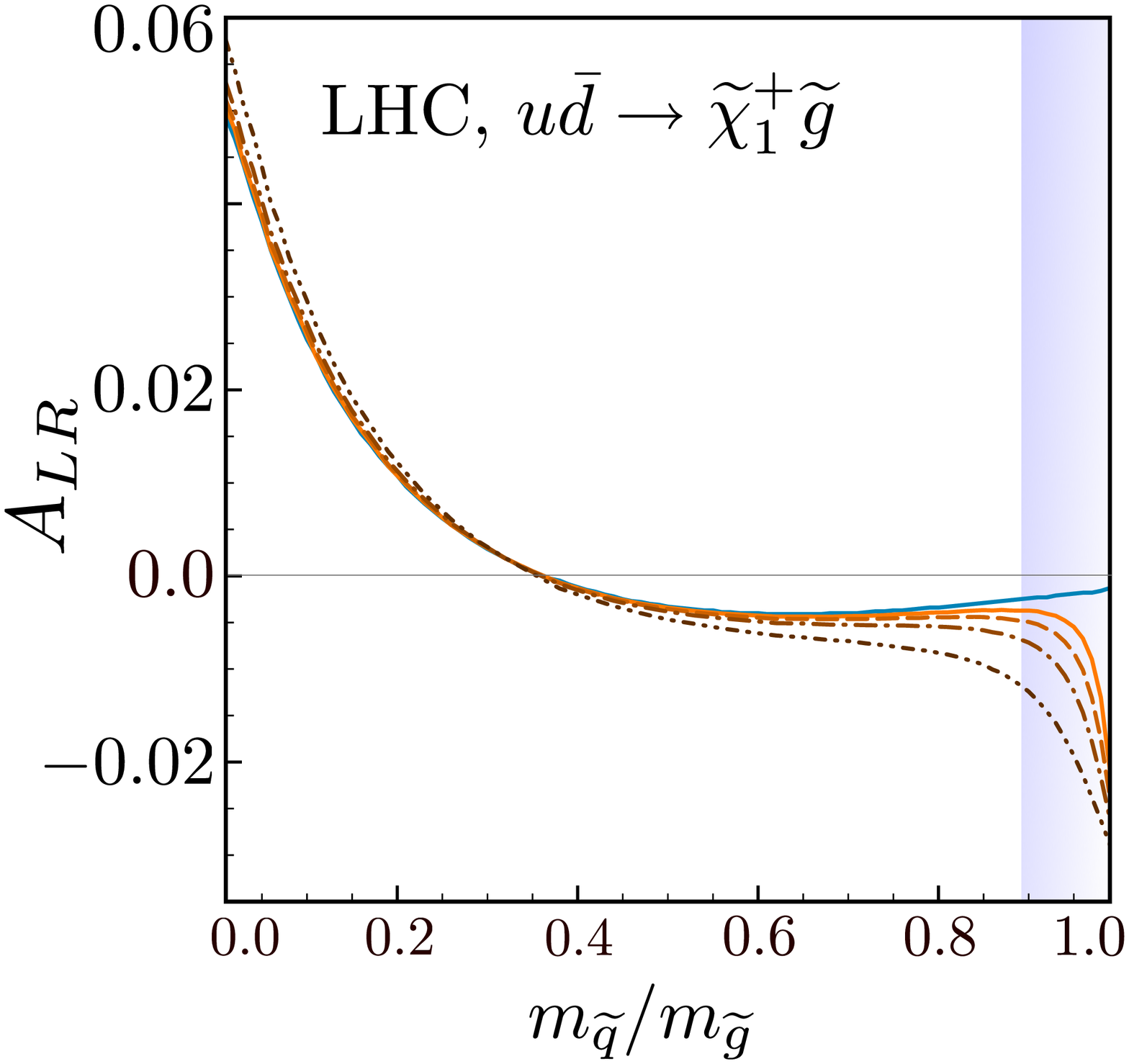}
  \end{center}
  \caption{Off-shell and interference effects for SUSY decay chains at
  the LHC. Left: shifts to the gluino BR as a function of
  $m_{\tilde{q}}/m_{\tilde{g}}$. Right: Decay asymmetry $(\tilde{g} \to
  \tilde{q}_L q) / (\tilde{g} \to \tilde{q}_R q)$; 
  from~\cite{berdine}. The shaded blue area covers all SPS points.}
  \label{fig:offshelleffects}   
\end{figure}

In Fig.~\ref{fig:offshelleffects} (taken from~\cite{berdine})
the effect of using full matrix elements are shown for the process
$u\bar d \to \tilde{\chi}_1^+ \tilde{g}$ with subsequent decays of the
gluino. The left plot shows the deviations of the effective branching
ratios for the gluino measured with the help of exclusive final states
as a function of the ratio of the squark over the gluino mass. As this
ratio approaches one, off-shell and interference effects give
corrections of the order of $20-40\,\%$ compared to the NWA. This
near-degenerate squark-gluino parameter region is quite generic, since
all SPS-standard candle points~\cite{SPS} lie in the blue band. So for
all of these points, there are large deviations from the NWA. The
right plot shows a chirality asymmetry, i.e. an asymmetry between the
decays $\tilde{g} \to \tilde{q}_L q$ and $\tilde{g} \to \tilde{q}_R
q$. 

In~\cite{catpiss++}, all of the off-shell and interference effects
discussed briefly above, are studied systematically for all kinds of
different SUSY decay chains and observables. Also, the size of these
effects as a function of the region of parameter space is
investigated.  


\section{Conclusions}

In conclusion, we validated the three multi-particle event generators
for MSSM processes, Madgraph, Sher\-pa, and WHIZARD, to produce correct
results. We studied the influence of using full matrix elements for
exclusive four- and six-particle final states in SUSY production
processes at the LHC and ILC compared to the narrow-width or
Breit-Wigner approximation. If there are several different diagram
classes that can become singly- or doubly-resonant, interferences of
these classes result in deviations from the NWA up to an order of
magnitude. Furthermore, off-shell effects are important for
nearly-degenerate mother-daughter constellations, as is typical in
SUSY decay chains (but also in all BSM scenarios with a conserved
parity and similar spectra). Especially, when cuts are mandatory to
disentangle signals and backgrounds, interference and off-shell
effects have to be studied by multi-particle event generators. 
 
\section{Acknowledgments} 
 
JR was partially supported by the Helmholtz-Gemein\-schaft under
Grant No. VH-NG-005 and the Bundes\-ministerium f\"ur Bildung und
Forschung, Germany, under Grant No. 05HA6VFB. Special thanks to David
Rainwater for his close collaboration and providing the
plots in Fig.~\ref{fig:offshelleffects}.

%

\begin{thebibliography}{999}
%
%


\bibitem{SPA}
  \url{http://spa.desy.de/spa}; 
  J.~A.~Aguilar-Saavedra {\it et al.},
  Eur.\ Phys.\ J.\  C {\bf 46}, 43 (2006).

\bibitem{trobens}
  T.~Robens, these proceedings; 
  W.~Kilian, J.~Reuter and T.~Robens,
  Eur.\ Phys.\ J.\  C {\bf 48}, 389 (2006);
  AIP Conf.\ Proc.\  {\bf 903}, 177 (2007)


\bibitem{omega}
  T. Ohl, \textit{O'Mega: An Optimizing Matrix Element Generator},
  hep-ph/0011243;
  M.~Moretti, T.~Ohl, J.~Reuter,
  hep-ph/0102195;
  J.~Reuter,
  arXiv:hep-th/0212154.

\bibitem{whizard}
  \texttt{http://whizard.event-generator.org};
  W.~Kilian. 
  LC-TOOL-2001-039, Jan 2001;
  W.~Kilian, T.~Ohl, J.~Reuter, 
  to appear in Comput.~Phys.~Commun., arXiv:0708.4233 [hep-ph].

\bibitem{sherpa}
  \url{http://www.sherpa-mc.de};
  F.~Krauss, R.~Kuhn and G.~Soff,
  JHEP {\bf 0202}, 044 (2002);
  T.~Gleisberg, S.~H\"oche, F.~Krauss, A.~Sch\"alicke, S.~Schumann and J.~C.~Winter,
  JHEP {\bf 0402}, 056 (2004).

\bibitem{madgraph}
  \url{http://madgraph.hep.uiuc.edu};
  H.~Murayama, I.~Watanabe and K.~Hagiwara, KEK-91-11;
  T.~Stelzer, F.~Long,
  Comput.{} Phys.{} Commun.{} \textbf{81} (1994) 357;
  F.~Maltoni and T.~Stelzer, JHEP {\bf 0302} (2003) 027.

\bibitem{SLHA}
  P.~Skands {\it et al.},
  JHEP {\bf 0407}, 036 (2004).

\bibitem{catpiss}
  K.~Hagiwara {\it et al.},
  Phys.\ Rev.\  D {\bf 73}, 055005 (2006);
  J.~Reuter {\it et al.},
  arXiv:hep-ph/0512012;
  J.~Reuter,
  arXiv:0709.0068 [hep-ph].

\bibitem{comparison}
  \url{http://whizard.event-generator.org/}\\
  \url{susy_comparison.html}

\bibitem{swi}
  T.~Ohl and J.~Reuter,
  Eur.\ Phys.\ J.\  C {\bf 30}, 525 (2003).

\bibitem{omwhiz_bsm}
  W.~Kilian and J.~Reuter,
  Phys.\ Rev.\ D {\bf 70} (2004) 015004;
  W.~Kilian, D.~Rainwater and J.~Reuter,
  Phys.\ Rev.\ D {\bf 71}, 015008 (2005);
  hep-ph/0507081;
  Phys.\ Rev.\  D {\bf 74}, 095003 (2006);
  M.~Beyer {\em et al.},
  Eur.\ Phys.\ J.\  C {\bf 48}, 353 (2006);
  W.~Kilian and J.~Reuter,
  hep-ph/0507099;
  J.~Reuter,
  arXiv:0708.4241 [hep-ph].
  T.~Ohl and J.~Reuter,
  Phys.\ Rev.\  D {\bf 70}, 076007 (2004);
  arXiv:hep-ph/0407337;
  W.~Kilian and J.~Reuter,
  Phys.\ Lett.\  B {\bf 642}, 81 (2006);
  F.~Deppisch, W.~Kilian, J.~Reuter, in preparation.

\bibitem{berdine}
  D.~Berdine, N.~Kauer and D.~Rainwater,
  to appear in Phys.\ Rev.\ Lett., arXiv:hep-ph/0703058.

\bibitem{SPS}
 B.~C.~Allanach {\it et al.},
  in {\it Proc. of the APS/DPF/DPB Summer Study on the Future of Particle Physics (Snowmass 2001) } ed. N.~Graf,
  Eur.\ Phys.\ J.\ C {\bf 25}, 113 (2002);
 N.~Ghodbane and H.~U.~Martyn,
  in {\it Proc. of the APS/DPF/DPB Summer Study on the Future of Particle Physics (Snowmass 2001) } ed. N.~Graf,
  arXiv:hep-ph/0201233.

\bibitem{catpiss++}
  A.~Alboteanu, J.~Alwall, W.~Kilian, T.~Plehn, D.~Rainwater, J.~Reuter,
  S.~Schumann, in preparation.

\end{thebibliography}
%

\end{document}